\documentclass[preprintnumbers,10pt,nofootinbib]{revtex4}
\pdfoutput=1

\usepackage{amsmath,latexsym,amssymb,amsfonts}
\usepackage[pdftex]{color,graphicx}
\usepackage{bm}

\begin{document}


\title{\textbf{Broken bridges: A counter-example of the ER=EPR conjecture}}

\author{
\textsc{Pisin Chen$^{a,b,c,d}$}\footnote{{\tt pisinchen{}@{}phys.ntu.edu.tw}},
\textsc{Chih-Hung Wu$^{a,b}$}\footnote{{\tt b02202007{}@{}ntu.edu.tw}}
and
\textsc{Dong-han Yeom$^{a}$}\footnote{{\tt innocent.yeom{}@{}gmail.com}}
}

\affiliation{
$^{a}$\small{Leung Center for Cosmology and Particle Astrophysics, National Taiwan University, Taipei 10617, Taiwan}\\
$^{b}$\small{Department of Physics, National Taiwan University, Taipei 10617, Taiwan}\\
$^{c}$\small{Graduate Institute of Astrophysics, National Taiwan University, Taipei 10617, Taiwan}\\
$^{d}$\small{Kavli Institute for Particle Astrophysics and Cosmology,
SLAC National Accelerator Laboratory, Stanford University, Stanford, California 94305, USA}
}

\begin{abstract}
In this paper, we provide a counter-example to the ER=EPR conjecture. In an anti-de Sitter space, we construct a pair of maximally entangled but separated black holes. Due to the vacuum decay of the anti-de Sitter background toward a deeper vacuum, these two parts can be trapped by bubbles. If these bubbles are reasonably large, then within the scrambling time, there should appear an Einstein-Rosen bridge between the two black holes. Now by tracing more details on the bubble dynamics, one can identify parameters such that one of the two bubbles either monotonically shrinks or expands. Because of the change of vacuum energy, one side of the black hole would evaporate completely. Due to the shrinking of the apparent horizon, a signal of one side of the Einstein-Rosen bridge can be viewed from the opposite side. We analytically and numerically demonstrate that within a reasonable semi-classical parameter regime, such process can happen. Bubbles are a non-perturbative effect, which is the crucial reason that allows the transmission of information between the two black holes through the Einstein-Rosen bridge, even though the probability is highly suppressed. Therefore, the ER=EPR conjecture cannot be generic in its present form and its validity maybe restricted.
\end{abstract}

\maketitle

\newpage

\tableofcontents


\section{Introduction}

The information loss problem \cite{Hawk} has been a great battlefield to clear the grounds for a future quantum theory of gravity. While there exist many candidate solutions to this problem \cite{Chen:2014jwq,Ong:2016iwi}, it is conceivable that the unitarity (at least for anti-de Sitter (AdS) space) would be preserved due to the AdS/CFT correspondence and the holographic principle \cite{Malda,Wit,GKP}.

In order to preserve the black hole unitarity while assuming the validity of both local quantum field theory and general relativity, the \textit{black hole complementarity principle} was conjectured \cite{LTU}. However, it was discovered that the original assumptions of black hole complementarity are seriously flawed \cite{Yeom:2008qw,Hong:2008mw,Yeom:2009zp}. In order to cure this inconsistency, Almheiri, Marolf, Polchinski and Sully (AMPS) \cite{AMPS,AMPSS} proposed that around the black hole horizon scale (after the Page time \cite{Page,Page2,Page3,Hwang:2016otg}), the equivalence principle must be violated; and they named such a region as the \textit{firewall}.

The key argument for the inconsistency of black hole complementarity can be described as follows \cite{MS}. First, due to unitarity, a particle from Hawking radiation \cite{Hawking:1974sw} (say, $B$) must be entangled (either after the Page time \cite{Page,Page2,Page3,Hwang:2016otg} or scrambling time \cite{Hayden:2007cs}, where the details depend on situations) with its counterpart $R_{B}$, where $R_{B}$ can be either the earlier part of Hawking radiation (for one-sided black holes) or the distilled information that is recovered at the boundary of the opposite side (for two-sided eternal black holes). Second, based on local quantum field theory, any given Hawking radiation $B$ should be entangled with its counterpart $A$ (which is inside the horizon). Then there appears a paradox since $B$ cannot be entangled with both $R_{B}$ and $A$ at the same time based on the monogamy of quantum entanglement. The AMPS resolution is to violate the equivalence principle and to introduce a high energy field gradient that breaks entanglement between $A$ and $B$, this is referred to as the firewall. 

The firewall conjecture may resolve the inconsistency of black hole complementarity, but there is no mechanism to really construct the firewall from the first principle, and hence it seems ad hoc. According to AMPS, the firewall conjecture would be the most conservative idea since it only changes the internal structure of the black hole. However, it has been later shown that if there exists a firewall, it can also affect the future infinity \cite{Hwang:2012nn,Kim:2013fv,COPSY}; it is therefore counter-argued that the firewall is not the most conservative proposal. But the question remains: is there any other way to circumvent the firewall?

One may think that the clarification of this problem should come from the holographic principle itself. One promising idea, although limited to (eternal) AdS black holes, is the \textit{ER=EPR conjecture} proposed by Maldacena and Susskind \cite{MS} (see also \cite{Suss,Suss2}). Before this conjecture was proposed, it has been already suggested \cite{Israel:1976ur} that if there is an Einstein-Rosen (ER) bridge system \cite{ER}, then both sides of the ER bridge are maximally entangled by Einstein-Podolsky-Rosen (EPR) pairs \cite{EPR}. This situation can be realized in the context of the AdS/CFT correspondence for two-sided AdS-Schwarzschild black holes. However, the essence of ER=EPR conjecture implies not only this but also its reverse argument; \textit{if there is a maximally entangled pair system, then their corresponding bulk geometry should have an Einstein-Rosen bridge between them} \cite{MS}. If this conjecture is true, then it may shed some light on the firewall proposal, because there can be \textit{connections} between $A$ and $R_{B}$ \textit{through the ER bridge} \cite{HMPS,MW,AC,VR}.

In this paper, we focus on the \textit{locality} of EPR pairs. As is well-known, there should be no transfer of information between EPR pairs. Then, based on the ER=EPR conjecture, the ER bridge should not be allowed to transmit any signal from one side to the other. However, we suggest a thought experiment, where the communication between the two sides does happen, which therefore violates the locality of EPR pairs. Since it is crucial for the ER=EPR conjecture to uphold the principle of locality, this suggests that the conjecture may not be true, at least not in its original form. On the other hand, if ER=EPR is proved to be correct in the end, then our thought experiment must have loopholes for some unknown reasons.

This paper is organized as follows. In SEC.~\ref{sec:EREPR}, we review the essence of the ER=EPR conjecture. In SEC.~\ref{sec:const}, we discuss the physical setting of our thought experiment, where this is divided by five steps (SECs.~\ref{sec:step1}, \ref{sec:step2}, \ref{sec:step3}, \ref{sec:step4}, and \ref{sec:step5}). Finally, in SEC.~\ref{sec:con}, we summarize this paper and discuss future perspectives.

\section{\label{sec:EREPR}Locality and the ER=EPR conjecture}

Spacetime locality means the impossibility of sending a signal faster than the speed of light. However, quantum mechanics sheds more light on the understanding of locality through the paradox of \textit{spooky action at a distance}, via, for example, the EPR entanglement. The name came from the fact that the EPR pair does not violate locality, but it indeed exhibits strange correlation between particles that are out of causal contact. The strange correlation of entanglement could be easily understood using a simple argument of Bell \cite{Bell}. Consider one entangles two spatially separated spin-$1/2$ particles into a maximal entangled pure state 
\begin{eqnarray}
| \psi \rangle = \frac{1}{\sqrt{2}} \left( | \uparrow \downarrow \;\rangle - | \downarrow \uparrow \;\rangle \right).
\end{eqnarray}
The expectation value of the total spin is zero, but the equal-time two-point functions,
\begin{eqnarray}
S_{AB}=\langle \psi | O_{A} O_{B} |\psi \rangle -\langle \psi | O_{A} |\psi \rangle\langle \psi | O_{B} |\psi \rangle,
\end{eqnarray}
would be non-zero. Since the two particles are spatially separated, by causality, this non-zero correlation should not be governed by the interaction between these two spinors. Instead, it indicates some non-trivial correlations in its density matrix. Actually for a pure state, any connected correlator with operators acting on causally disconnected degrees of freedom would provide a good quantitative description of the EPR entanglement. However, we should emphasize that the correlation of entanglement cannot be used to transmit messages superluminally.

The ER bridge in general relativity also shares similar correlations between distant black holes. The ER bridge is non-traversable in the sense that one observer outside the event horizon cannot be connected by a time-like curve to an outside observer of the other side of the ER bridge. The only thing the two observers can do is to meet inside the event horizon of the ER bridge, rather than communicate through the bridge. Note that the topological censorship theorems forbid traversable wormholes under some reasonable energy conditions \cite{FW,FSW,GSWW}, which maybe also true for eternal black holes \cite{Kelly:2014mra}.

According to the ER=EPR conjecture, the ER bridge is a dual system to the EPR entangled pairs. Prior to the ER=EPR conjecture, one already knew that in the context of the AdS/CFT correspondence, the Hartle-Hawking states \cite{Hartle:1976tp} could serve as a bulk ground state for the AdS ER bridge
\begin{eqnarray}
|\psi \rangle= \frac{1}{Z} \sum_{i} e^{-\frac{\beta E_i}{2}} |i^{*}\rangle_{L} |i\rangle_{R},
\end{eqnarray}
where $Z$ is the partition function, $\beta$ is the inverse temperature, and $i$ labels eigenstates of the Hamiltonian in the left and right exteriors, respectively \cite{Malda2}. Therefore, the ER bridge generates two entangled regions by Hawking particles. Adding to this, the proposal of building up spacetime through quantum entanglement \cite{VR} has motivated the ER=EPR conjecture. This further argues that for a given EPR pair, there should appear an ER bridge. The most radical form of the conjecture is that a Bell pair of the Hawking quanta must be connected with the black hole interior via Planckian wormholes to encode the entanglement \cite{MS}. After the Page time, if the radiated Hawking quanta can be collapsed into another black hole, then all the microscopic wormholes would combine into a single macroscopic ER bridge. We explicitly quote from the original paper about an unfeasible logical jump \cite{MS}:
\begin{quote}
\textit{Finally let us make a jump. Suppose that we take a large number of particles, entangled
into separate Bell pairs, and separate them in the same way as the mini-black holes. When
we collapse each side to form two distant black holes, the two black holes will be entangled.
We make the conjecture that they will also be connected by an Einstein-Rosen bridge. In
fact we go even further and claim that even for an entangled pair of particles, in a quantum
theory of gravity there must be a Planckian bridge between them, albeit a very quantum mechanical
bridge which probably cannot be described by classical geometry.}
\end{quote}
What we are really constructing is not to falsify the AdS/CFT duality, but to reveal the inconsistency of this unfeasible link. 

In order to invoke the ER=EPR conjecture, the ER bridge should share some properties with the EPR pairs, especially the locality. In terms of the ER bridge, one requires that the ER bridge should not be traversable. However, if we change physical conditions of asymptotic regions, then the horizon can in principle shrink. In this case, the null energy condition can be violated \cite{Visser:1996iw,Visser:1996iv} and the situation will be completely different. If this is the case, then the ER=EPR conjecture would be in doubt. Now we go to the physical construction of the setting that can reveal this situation, showing that there is a tension between the ER=EPR conjecture and quantum mechanics.

\section{\label{sec:const}Construction of the setting}

In this section, we consider a thought experiment based on general relativity, local quantum field theory, and the ER=EPR conjecture. In Step 1 (SEC.~\ref{sec:step1}), we generate a maximally entangled pair system (say $L$ and $R$) in an AdS background. In Step 2 (SEC.~\ref{sec:step2}), we further assume that, due to the vacuum decay process based on quantum field theory, each part of the pair is trapped by true vacuum bubbles, where one is expanding and the other is contracting. Soon after these are trapped by bubbles, by scrambling one side, (assuming the ER=EPR conjecture) we can induce the ER bridge that connects $L$ and $R$. In Step 3 (SEC.~\ref{sec:step3}), as we assumed, the contracting bubble completely collapses into one of the black holes, say, $R$. This will change the boundary condition of $R$ side and can induce the shrink of the horizon. In Step 4 (SEC.~\ref{sec:step4}), we provide the details of the apparent horizon behaviors. We give some physical conditions within the semi-classical quantum field theory that induces the situation that the true event horizon is way inside the putative event horizon (that would be the event horizon if the apparent horizon does not shrink). After the evaporation is completed, in Step 5 (SEC.~\ref{sec:step5}), the ER bridge will be broken. In this causal structure, a signal from the left side can be transferred to the right side. This completes our thought experiment.

From now, we show the technical details of this thought experiment.

\subsection{\label{sec:step1}Step 1: Generation of an entangled system}

We assume Einstein gravity with a number of scalar fields, i.e.,
\begin{eqnarray}
S = \int \sqrt{-g} dx^{4} \left[ \frac{\mathcal{R}}{16\pi} - \frac{1}{2} \left( \nabla \phi \right)^{2} - U(\phi) - \sum_{i=1}^{N} \frac{1}{2} \left( \nabla f_{i} \right)^{2} \right],
\end{eqnarray}
where $\mathcal{R}$ is the Ricci scalar, $\phi$ is a scalar field with a potential $U(\phi)$, and $f_{i}$ are $N$ number of massless scalar fields (that will contribute to Hawking radiation). We further assume that $U(\phi)$ has at least two minima $\Lambda_{+}$ and $\Lambda_{-}$, with $\Lambda_{-} < \Lambda_{+} < 0$. Therefore, $\Lambda_{-}$ is the true vacuum and $\Lambda_{+}$ is the false vacuum.

\begin{figure}
\begin{center}
\includegraphics[scale=0.75]{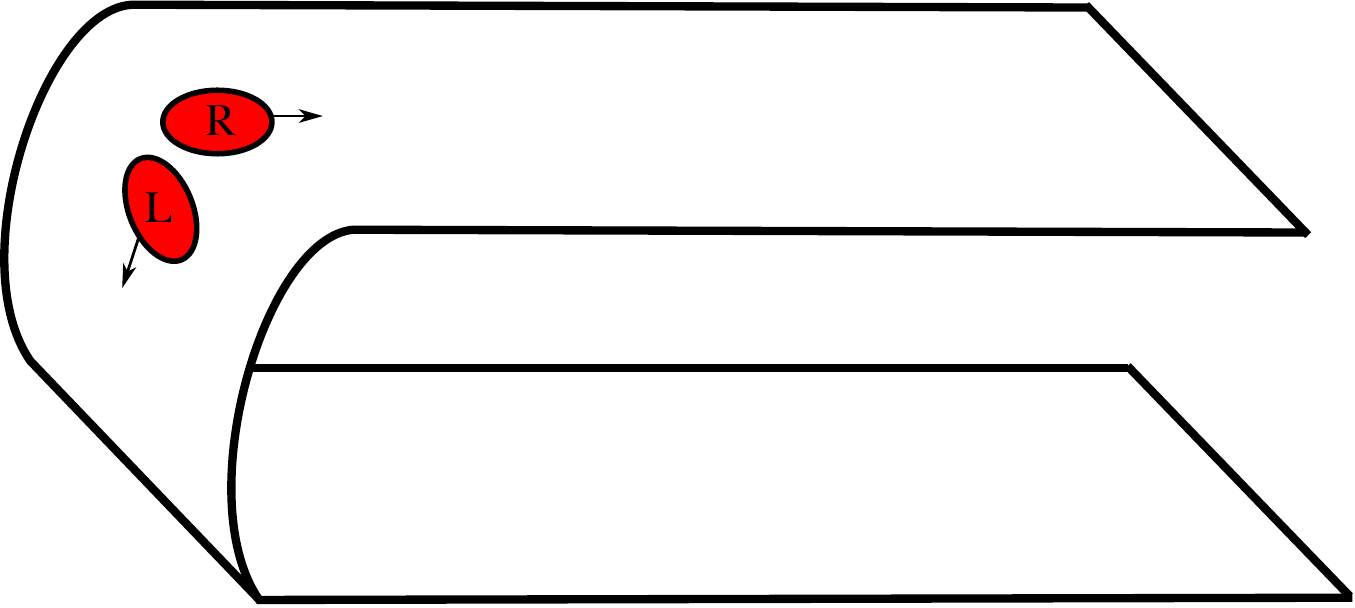}
\caption{\label{fig:step1}Generation of an entangled system.}
\end{center}
\end{figure}

Let us consider a background with cosmological constant $\Lambda_{+}$, where $|\Lambda_{+}| = 3/8\pi \ell_{+}^{2}$. In addition, we construct a situation such that within this AdS space, two entangled systems are generated and separated by a long distance (FIG.~\ref{fig:step1}). Let us say that one is $L$ and the other is $R$.

We can view this construction as having a resource of entangled systems, which is prepared in advance. For example, we can think of a third observer who created a collection of such entangled systems and sent half of them to $L$ and $R$, respectively. 

\subsection{\label{sec:step2}Step 2: Trapping by bubbles and formation of the Einstein-Rosen bridge}

\begin{figure}
\begin{center}
\includegraphics[scale=0.6]{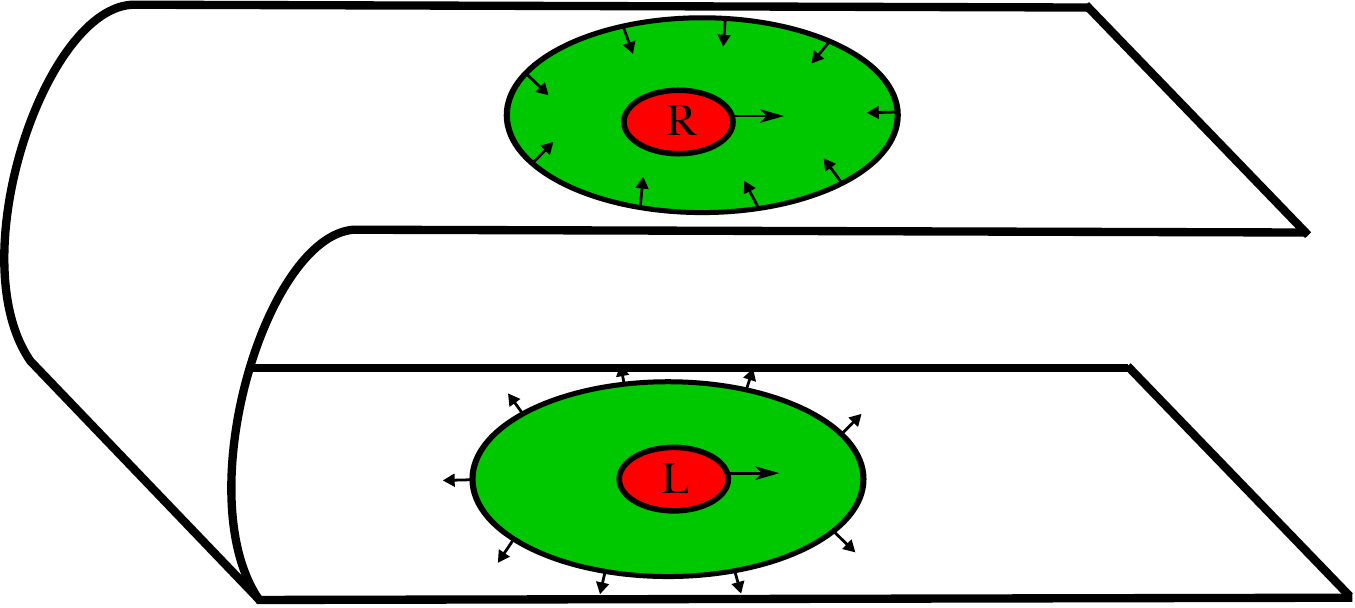}
\includegraphics[scale=0.6]{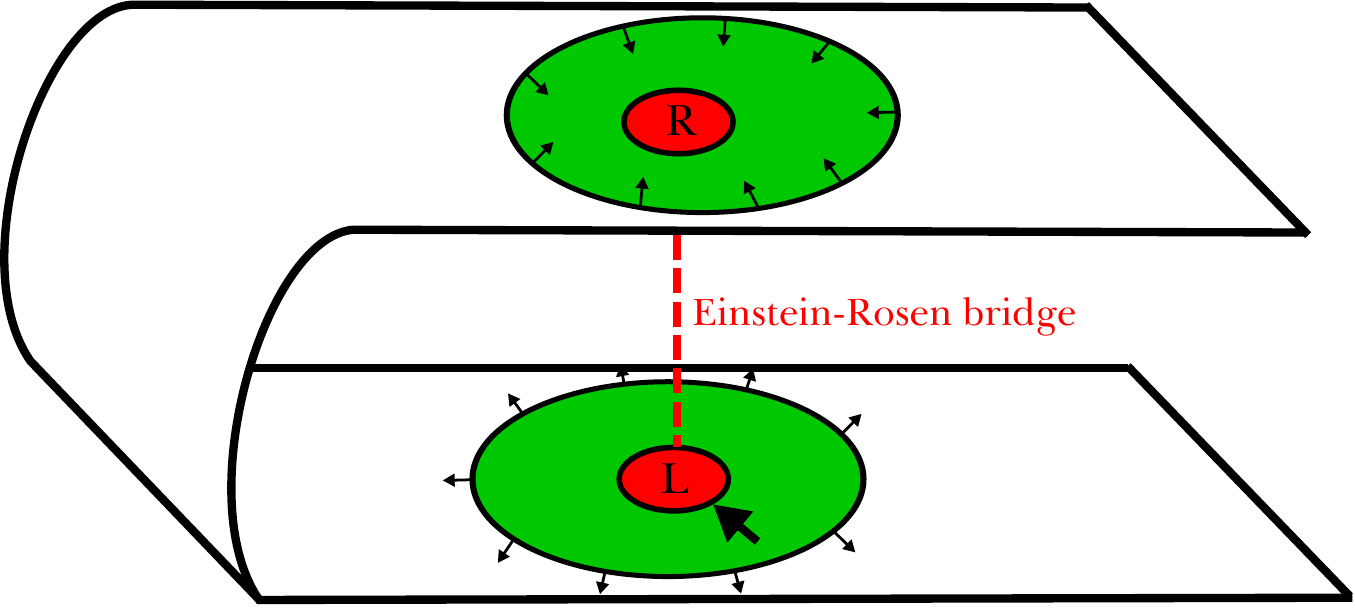}
\caption{\label{fig:step2}Left: Each of the two nucleated bubbles, one expanding and the other contracting, traps an entangled region. Right: When bubbles are still large enough, by scrambling one side, there appears an ER bridge between two entangled systems.}
\end{center}
\end{figure}

Second, we assume that in this AdS space, two true vacuum bubbles with vacuum energy $\Lambda_{-}$ are nucleated. While there should exist the most preferred size of the bubble (e.g., \cite{Coleman:1980aw}), in principle any size and initial condition of the shell would suffice our purpose. Hence, we assume that at the beginning the two bubbles are so large that they separately trap the two entangled systems $L$ and $R$. In addition, it is conceivable that there exist solutions such that one bubble is contracting and the other is expanding. These expanding and contracting bubbles can be well described by the spherically symmetric thin-shell dynamics (see more details in the next subsection, left of FIG.~\ref{fig:step2}).

Inside each bubble, we assume the process of scrambling described by the random unitary matrix chosen from the group-invariant Haar measure. Then, as the system is scrambled, the ER bridge is induced according to the ER=EPR conjecture (right of FIG.~\ref{fig:step2}).

One interesting comment is that this situation is very similar to the lab experiment that was introduced by Maldacena and Susskind \cite{MS}. Here, we explicitly quote one paragraph from their paper:
\begin{quote}
\textit{Alice has prepared a pair of identical spherical shells, $S_L$ and $S_R$, made of a special material that supports a large (but finite) $N$ CFT with a gravity dual. Note that Alice is not a creature of the bulk. She lives in ordinary space outside the shells; that is, in the laboratory. We will assume that Alice has complete control over the shells. In particular, by applying a variety of external fields she can manipulate the Hamiltonians of the shells in an arbitrary manner. She can also distill quantum information from a shell, and transport it from one shell to another; all of this as rapidly and accurately as she likes. As for the constraints of special relativity, we may assume that the velocity of propagation on the shells is much slower than the speed of light in the laboratory, so that Alice is unconstrained by any speed limit. She cannot however go backward in time.}
\end{quote}
In their paper, they introduced two shells where inside the shell is AdS and two shells are entangled with each other. Each shell is large enough and hence the internal geometry of the shell can be approximated by an AdS black hole. Due to the ER=EPR conjecture, two shells are connected by the ER bridge and they performed thought experiments within this system. Their setting is almost the same as what we have introduced; the only difference is that we are now giving \textit{dynamics} to shells, where one side is expanding (hence, we do not need to care) and the other side is contracting that we have to take a closer look into the details.

\subsection{\label{sec:step3}Step 3: Collapsing of the bubble for one side}

In order to show the existence of such a collapsing and expanding shell, we can assume the spherical symmetry for simplicity. After the ER bridge formation, we can assume the geometry of $L$ and $R$ as a black hole with a given mass. Now we assume the system: there is a shell at $r(t)$ and the outside $(+)$ and inside $(-)$ the shell satisfy
\begin{eqnarray}
ds^{2} = - f_{\pm}(R) dT^{2} + \frac{1}{f_{\pm}(R)} dR^{2} + R^{2} d\Omega^{2},
\end{eqnarray}
where
\begin{eqnarray}
f_{\pm} = 1 - \frac{2M_{\pm}}{R} + \frac{R^{2}}{\ell_{\pm}^{2}},
\end{eqnarray}
$\ell_{\pm}^{2} = 3 / 8 \pi |\Lambda_{\pm}|$, and $M_{\pm}$ denote mass parameters for outside and inside the shell.

For future applications, we impose the following conditions:
\begin{itemize}
\item[--] 1. Inside the shell, the past eternity condition is satisfied:
\begin{eqnarray}\label{eq:HPt}
T_{H} = \frac{\ell_{-}^{2} + 3 r_{h}^{2}}{4\pi \ell_{-}^{2} r_{h}} > \frac{1}{\pi \ell_{-}},
\end{eqnarray}
where $T_{H}$ is the Hawking temperature of the black hole. This is the condition that the black hole is preferred than thermal radiation \cite{Hawking:1982dh}.
\item[--] 2. Outside the shell is approximately Minkowski (false vacuum): $\ell_{+} \ggg \ell_{-}$. Hence, after the shell collapses, the black hole starts to shrink due to Hawking radiation.
\item[--] 3. The shell satisfies the null energy condition, i.e., the tension of the shell $\sigma$ is positive and $M_{+} > M_{-}$.
\end{itemize}
Note that if the shell is made by a scalar field, then the tension $\sigma$ can be assumed as a constant \cite{Blau:1986cw}.

This implements a situation of the false vacuum decay. Regarding this, one may ask about the instability of AdS. It was pointed out in \cite{Horowitz:2007pr, Harlow:2010az} that if the decay rate per unit volume is nonzero, the causal structure of the AdS space do not allow the false vacuum to exist longer than a time of order the AdS scale $\sim \ell_{+}$, nor it would convert into a cosmological singularity. However, this fact is no more harmful for our construction since we will assume $\ell_{+} \ggg 1$, making the time scale of the AdS space arbitrarily large.

The shell should satisfy the following equation of motion \cite{Israel:1966rt}:
\begin{eqnarray}
\epsilon_{-} \sqrt{\dot{r}^{2}+f_{-}(r)} - \epsilon_{+} \sqrt{\dot{r}^{2}+f_{+}(r)} = 4\pi r \sigma,
\end{eqnarray}
where $\epsilon_{\pm}$ denote the sign of the outward normal directions and these are proportional to the extrinsic curvatures:
\begin{eqnarray}
\beta_{+}(r) &\equiv&
 \frac{f_{-}(r)-f_{+}(r)-16\pi^{2} \sigma^{2} r^{2}}{8 \pi \sigma r}
 = \epsilon_{+} \sqrt{\dot{r}^{2}+f_{+}(r)}\,,
\\ \label{eq:ec2}
\beta_{-}(r) &\equiv&
 \frac{f_{-}(r)-f_{+}(r)+16\pi^{2} \sigma^{2} r^{2}}{8 \pi \sigma r}
 = \epsilon_{-} \sqrt{\dot{r}^{2}+f_{-}(r)}\,.
\end{eqnarray}
The equation of motion of the shell can be reduced as
\begin{eqnarray}
\dot{r}^{2} + V(r) &=& 0,\,
\label{eq:form}
\\
V(r) &=& f_{+}(r)- \frac{\left(f_{-}(r)-f_{+}(r)
-16\pi^{2} \sigma^{2} r^{2}\right)^{2}}{64 \pi^{2} \sigma^{2} r^{2}}.
\end{eqnarray}
In this effective potential $V(r)$, the physically allowed solution is for $V(r) < 0$.

Now we further impose conditions on the shell dynamics:
\begin{itemize}
\item[--] 1. The shell allows an asymmetric collapsing solution (starting from infinity, monotonously collapses to the singularity). In other words, $V(r) < 0$ for all $r$.
\item[--] 2. The shell starts from the right side of the Penrose diagram, i.e., $\beta_{\pm} > 0$ for $r \rightarrow \infty$. This is equivalent to the condition: $\ell_{-}^{-2} - \ell_{+}^{-2} - 16 \pi^{2} \sigma^{2} > 0$.
\end{itemize}
One can choose a parameter space that satisfies all the constraints by tuning proper $\ell_{-}$ and $\sigma$.

\subsubsection{Conditions for $V(r) < 0$}

After simple calculations, we can reduce the junction equation by a simpler formula:
\begin{eqnarray}\label{eq:form2}
V(r) = 1 - \left(-\frac{1}{\ell_{+}^{2}} + \frac{\mathcal{B}^{2}}{64 \pi^{2} \sigma^{2}}\right) r^{2} - 2M_{+}\left( 1 + \frac{\Delta}{M_{+}} \frac{\mathcal{B}}{32 \pi^{2} \sigma^{2}} \right) \frac{1}{r}  - \frac{\Delta^{2}}{16 \pi^{2} \sigma^{2}} \frac{1}{r^{4}},
\end{eqnarray}
where
\begin{eqnarray}
\mathcal{B} &\equiv& \frac{1}{\ell_{-}^{2}} - \frac{1}{\ell_{+}^{2}} - 16 \pi^{2} \sigma^{2},\\
\Delta &\equiv& M_{+} - M_{-}.
\end{eqnarray}

For simplicity, we can introduce the following new parametrization \cite{Aguirre:2005xs,Aguirre:2005nt,Yeom:2016qec}:
\begin{eqnarray}
z &\equiv& \left( \frac{\sqrt{\mathcal{B}^{2} - 64\pi^{2}\sigma^{2}/\ell_{+}^{2}}}{2 \Delta} \right)^{1/3} r,\\
\eta &\equiv& \left( \frac{\sqrt{\mathcal{B}^{2} - 64\pi^{2}\sigma^{2}/\ell_{+}^{2}}}{8\pi\sigma} \right) t,\\
L_{\pm} &\equiv& 4 \pi\sigma \ell_{\pm},\\
\tilde{M}_{\pm} &\equiv& \frac{M_{\pm}}{\pi \sigma}.
\end{eqnarray}
Then one can simplify the equation of motion:
\begin{eqnarray}\label{eq:form}
\left(\frac{dz}{d\eta}\right)^{2} + U(z) = - \mathcal{E},
\end{eqnarray}
where
\begin{eqnarray}
U(z) &=& - \left( z^{2} + \frac{\mathcal{C}}{z} + \frac{1}{z^{4}} \right),\\
\mathcal{C} &=& \frac{2 (L_{-}^{-2} - L_{+}^{-2} - 1 + 2 \tilde{M}_{+}/\tilde{\Delta})}{\sqrt{(L_{-}^{-2} - L_{+}^{-2} - 1)^{2} - 4 L_{+}^{-2}}},\\
\mathcal{E}^{3} &=& \frac{1}{\tilde{\Delta}^{2} \pi^{4} \sigma^{4}} \frac{1}{\left( (L_{-}^{-2} - L_{+}^{-2} - 1)^{2} - 4 L_{+}^{-2} \right)^{2}}.
\end{eqnarray}

\begin{figure}
\begin{center}
\includegraphics[scale=0.6]{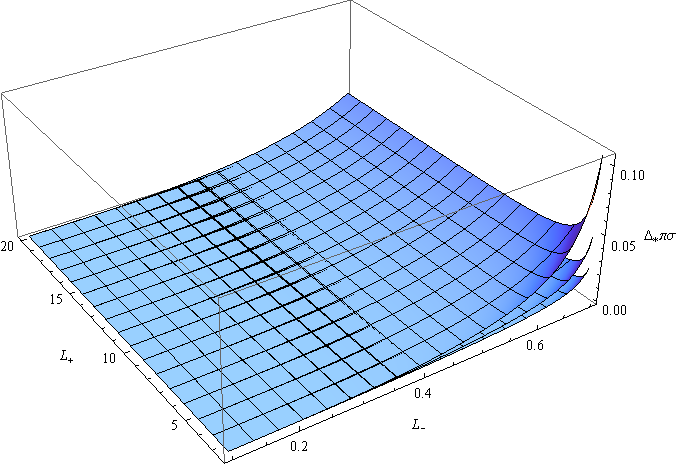}
\caption{\label{fig:Deltastar}$\Delta_{*}\pi \sigma$ for a given $\Delta/M_{+} = 0.1$, $0.2$, and $0.5$ (bottom to top) as a function of $L_{+}$ and $L_{-}$, while we are searching the region satisfying $L_{-}^{-2} - L_{+}^{-2} - 1 > 0$.}
\end{center}
\end{figure}

The condition for the existence of the extreme limit is as follows: there exists $z_{0}$ such that $U'(z_{0}) = 0$ and $\left| U(z_{0}) \right| = \mathcal{E}$. Then $z_{0}$ should satisfy
\begin{eqnarray}
z_{0}^{6} - \frac{\mathcal{C}}{2} z_{0}^{3} - 2 = 0.
\end{eqnarray}
The physical solution is
\begin{eqnarray}
z_{0} = \left( \frac{\mathcal{C} + \sqrt{\mathcal{C}^{2} + 32}}{4} \right)^{1/3}.
\end{eqnarray}
Therefore, if
\begin{eqnarray}
M_{+} - M_{-} > \Delta_{*} \equiv \frac{1}{\pi\sigma} \frac{\left( (\mathcal{C} + \sqrt{\mathcal{C}^{2} + 32})/4 \right)^{2}}{(L_{-}^{-2} - L_{+}^{-2} - 1)^{2} - 4 L_{+}^{-2}} \left[3+ \frac{3\mathcal{C}}{2} \left(\frac{\mathcal{C} + \sqrt{\mathcal{C}^{2} + 32}}{4}\right)\right]^{-3/2},
\end{eqnarray}
then $V(r) < 0$ is satisfied (FIG.~\ref{fig:Deltastar}). Especially, in the $L_{+} \rightarrow \infty$ and $\Delta / M_{+} \ll 1$ limit,
\begin{eqnarray}
\Delta_{*} \simeq \frac{1}{6\sqrt{3} \pi\sigma \left(L_{-}^{-2} - 1\right)} \frac{\Delta}{M_{+}} + \mathcal{O}\left(\frac{\Delta^{2}}{M_{+}^{2}}\right).
\end{eqnarray}
Hence, if $\Delta / M_{+} \ll 1$, as long as
\begin{eqnarray}
M_{+} \gtrsim \frac{1}{6\sqrt{3} \pi\sigma \left(L_{-}^{-2} - 1\right)}
\end{eqnarray}
is satisfied, one can obtain the condition $V(r) < 0$ for all $r$.

\subsubsection{Apparent horizon difference}

\begin{figure}
\begin{center}
\includegraphics[scale=0.75]{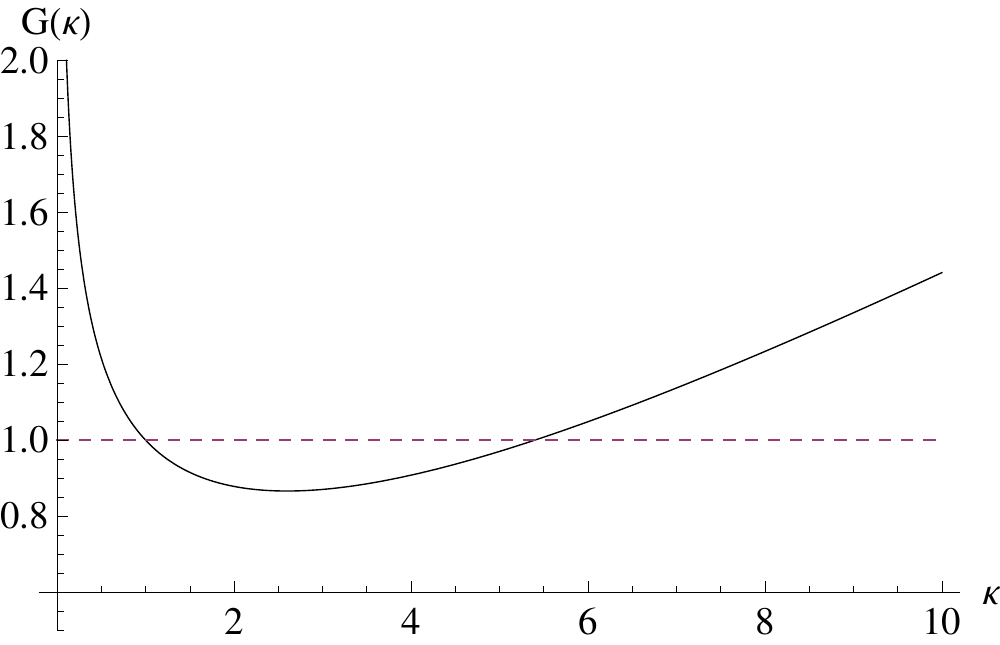}
\includegraphics[scale=0.75]{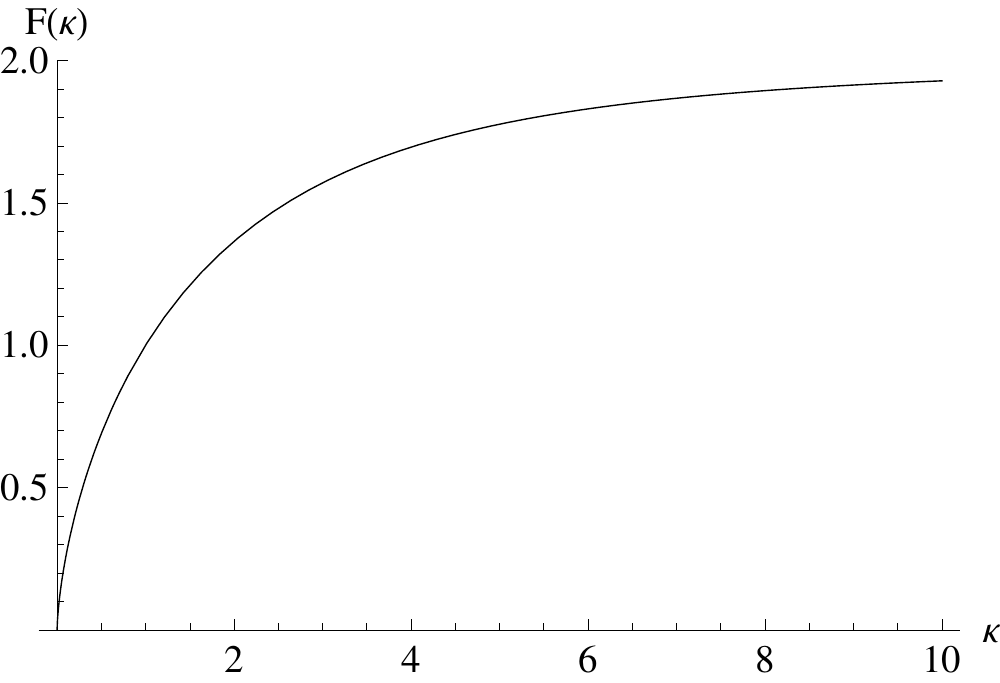}
\caption{\label{fig:Gg}Left: $G(\kappa)$ as a function of $\kappa$. Hence, if $\kappa \gtrsim 5.5$, then it can satisfy the past eternity condition. Right: $F(\kappa)$ as a function of $\kappa$.}
\end{center}
\end{figure}

The horizon before the collapse $r_{h}$ is defined by $f_{-}(r_{h}) = 0$ and the solution is
\begin{eqnarray}
r_{h} = \frac{\left(9 \ell_{-}^{2} M_{-} + \sqrt{3}\sqrt{\ell_{-}^{6} + 27 \ell_{-}^{4} M_{-}^{2}}\right)^{1/3}}{3^{2/3}} - \frac{\ell_{-}^{2}}{3^{1/3} \left( 9 \ell_{-}^{2} M_{-} + \sqrt{3} \sqrt{\ell_{-}^{6} + 27 \ell_{-}^{4} M_{-}^{2}} \right)^{1/3}}.
\end{eqnarray}
If $\ell_{-} = \kappa M_{-}$, then
\begin{eqnarray}
\frac{r_{h}}{M_{-}} = F(\kappa) \equiv \frac{\left(9 \kappa^{2} + \sqrt{3}\sqrt{\kappa^{6} + 27 \kappa^{4}}\right)^{1/3}}{3^{2/3}} - \frac{\kappa^{2}}{3^{1/3} \left( 9 \kappa^{2} + \sqrt{3} \sqrt{\kappa^{6} + 27 \kappa^{4}} \right)^{1/3}}.
\end{eqnarray}
Then the condition for the past eternity (Eq.~(\ref{eq:HPt})) is
\begin{eqnarray}
G(\kappa) \equiv \frac{\kappa^{2} + 3 F^{2}(\kappa)}{4 \kappa F(\kappa)} > 1.
\end{eqnarray}
FIG.~\ref{fig:Gg} shows numerical behaviors of $G(\kappa)$ and $F(\kappa)$.

The apparent horizon after the collapse is $r_{\mathrm{ApH}} = 2M_{+}$ (assuming $\ell_{+} \gg M_{+}$). Therefore, their radial differences are
\begin{eqnarray}
\frac{|r_{h} - r_{\mathrm{ApH}}|}{2M_{+}} \equiv \frac{\delta r}{2M_{+}},
\end{eqnarray}
where
\begin{eqnarray}
r_{h} = F(\kappa) M_{-}
\end{eqnarray}
and $F(\kappa)$ for the allowed region is approximately $F(\kappa) \sim 2$. In more details,
\begin{eqnarray}
\delta r = 2\Delta + \frac{8 M_{-}}{\kappa^{2}} + \mathcal{O}\left(\kappa^{-4}\right).
\end{eqnarray}

\begin{figure}
\begin{center}
\includegraphics[scale=0.6]{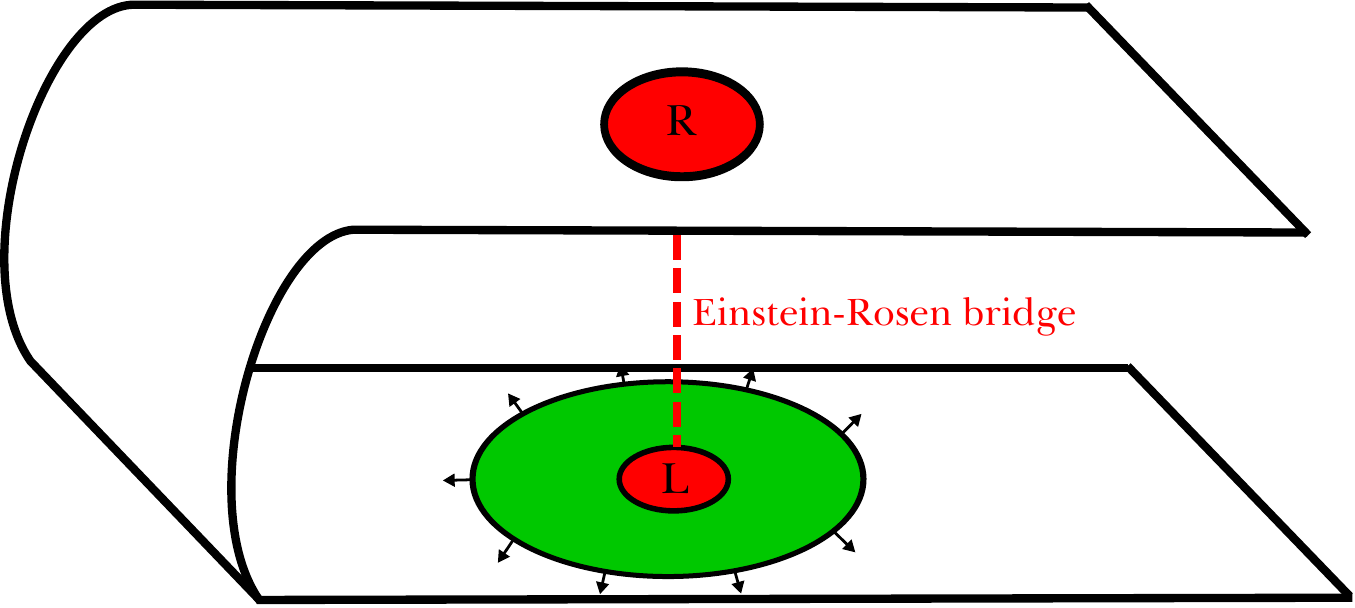}
\includegraphics[scale=0.6]{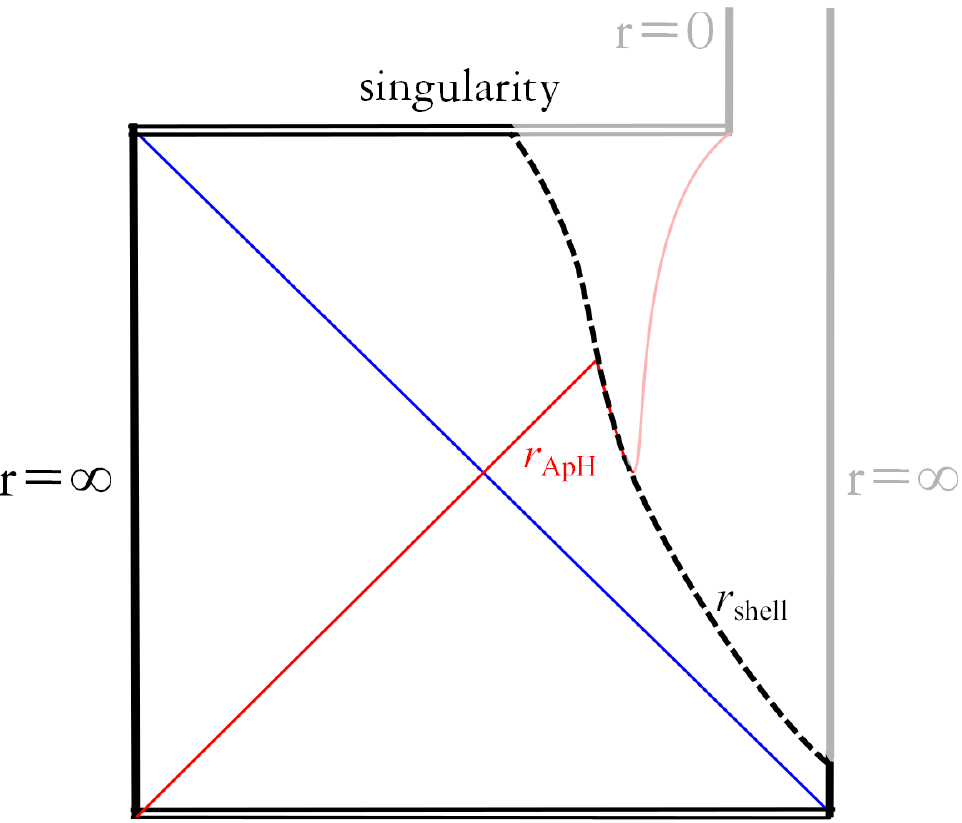}
\caption{\label{fig:step3}Due to the collapsing of the shell, the black hole $R$ slightly increased.}
\end{center}
\end{figure}

\subsubsection{Causal structures}

Now we can find such an example that it allows the past eternity condition as well as collapsing or expanding solution. Hence, we can assume that for the left side, the bubble is expanding (and hence, it is reasonable to assume that such an expanding bubble is in the far past and can be ignored from our sight), while for the right side, the bubble should shrink and collapse into the black hole. Then for the right side, we need to describe here as a black hole within the cosmological constant $\Lambda_{+}$. In addition, after the collapse, the black hole mass should be increased from $M_{-}$ to $M_{+}$; the horizon should be increased from $r_{h}$ to $r_{\mathrm{ApH}}$ (left of FIG.~\ref{fig:step3}). Finally, the causal structure should look like the right of FIG.~\ref{fig:step3}, while this is only true for the geometry of the $f_{-}$ (inside the shell) and the shell. For the causal structure of the $f_{+}$ part, we need more considerations.

\subsection{\label{sec:step4}Step 4: Evaporation of the black hole}

\begin{figure}
\begin{center}
\includegraphics[scale=0.6]{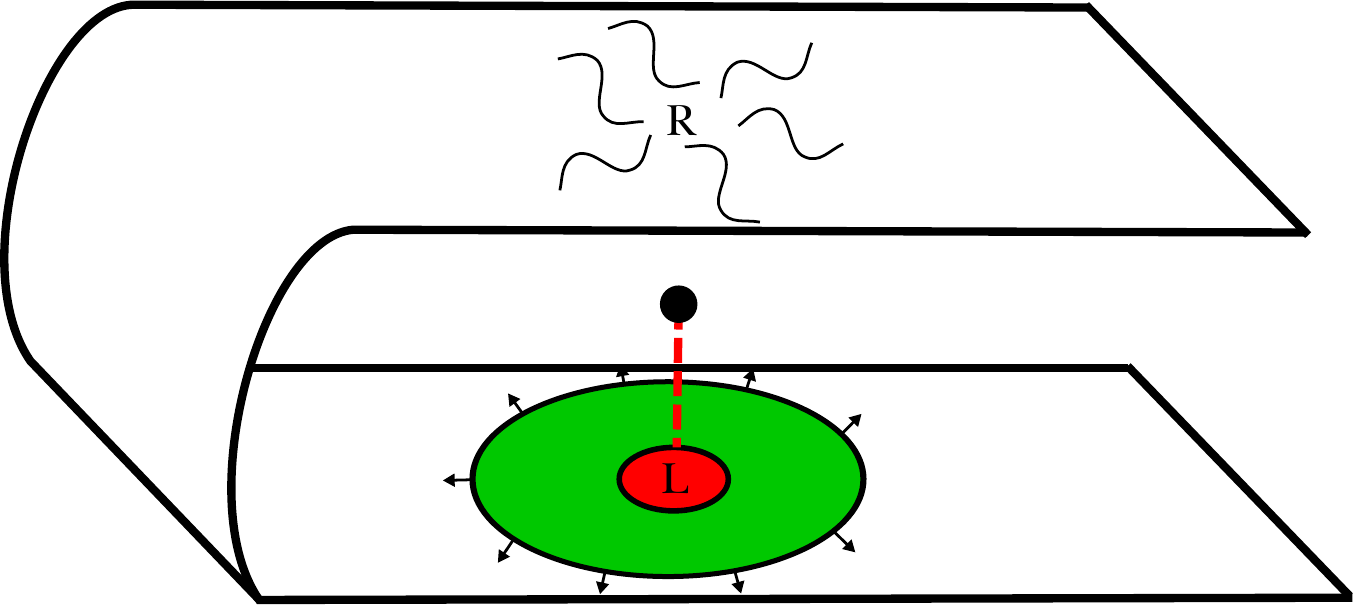}
\caption{\label{fig:step4}After the evaporation, the black hole $R$ disappears and the ER bridge is disconnected.}
\end{center}
\end{figure}

Since the cosmological constant of $R$ is changed, we can induce evaporation. We calculate more details of the evaporation and apparent horizon dynamics. Eventually, the requirements for the communication between $L$ and $R$ through the ER bridge are satisfied.

\subsubsection{Adiabatic process} Even though the black hole is stable for $f_{-}$ region, by changing the cosmological constant, the black hole for the right side can evaporate (FIG.~\ref{fig:step4}). In order to realize this process, by assuming $\ell_{+} \gg M_{+}$, we can use the Vaidya metric approximation \cite{Vaidya:1951zz}:
\begin{eqnarray}
ds^{2} = - \left( 1 - \frac{2M(v)}{r} \right) dv^{2} + 2 dv dr + r^{2} d\Omega^{2},
\end{eqnarray}
where $M(v = 0) = M_{+}$. For each out-going null geodesics, it should satisfy
\begin{eqnarray}
\frac{dr}{dv} = \frac{1}{2} \left(1 - \frac{2M}{r} \right),
\end{eqnarray}
and for the usual adiabatic process \cite{Hawking:1974sw},
\begin{eqnarray}
\frac{dM}{dv} = - \frac{\alpha N}{M^{2}},
\end{eqnarray}
where $\alpha$ is a constant that is proportional to the Stefan-Boltzmann constant and $N$ is the number of matter fields that contribute to Hawking radiation. By solving these two equations, we can obtain the dynamical behaviors of $r(v)$ and $M(v)$. Especially, if $\ell_{+} \gg M_{+}$, then the apparent horizon is $r_{\mathrm{ApH}} = 2M$ and the event horizon $r_{\mathrm{EH}}$ is the out-going null geodesic that satisfies
\begin{eqnarray}
r_{\mathrm{EH}}(v_{\mathrm{max}}) = 0,
\end{eqnarray}
where $v_{\mathrm{max}}$ is defined by $M(v_{\mathrm{max}}) = 0$ (i.e., the end point of the evaporation).

\begin{figure}
\begin{center}
\includegraphics[scale=0.75]{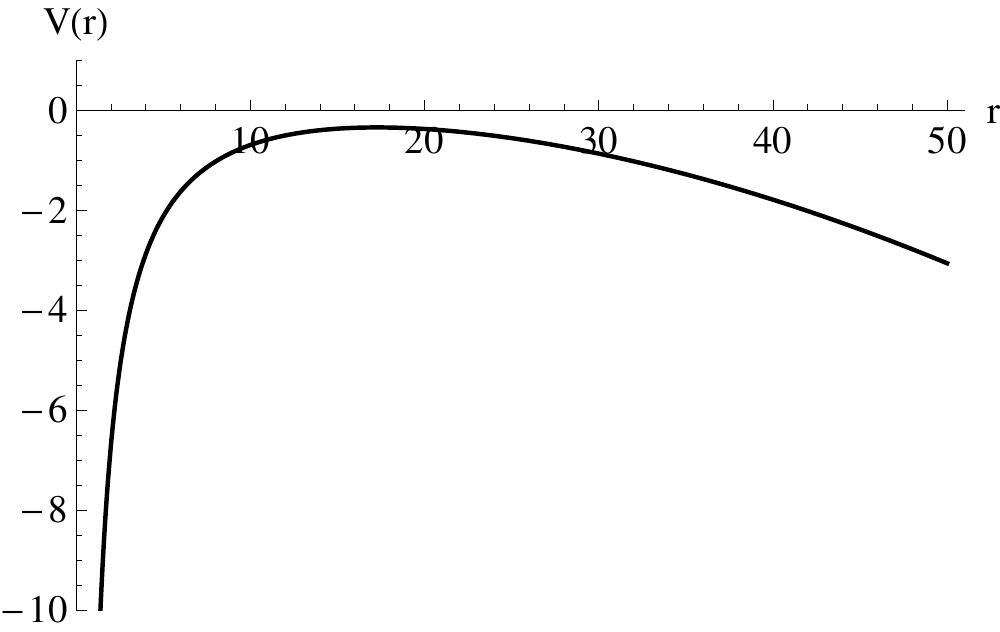}
\includegraphics[scale=0.75]{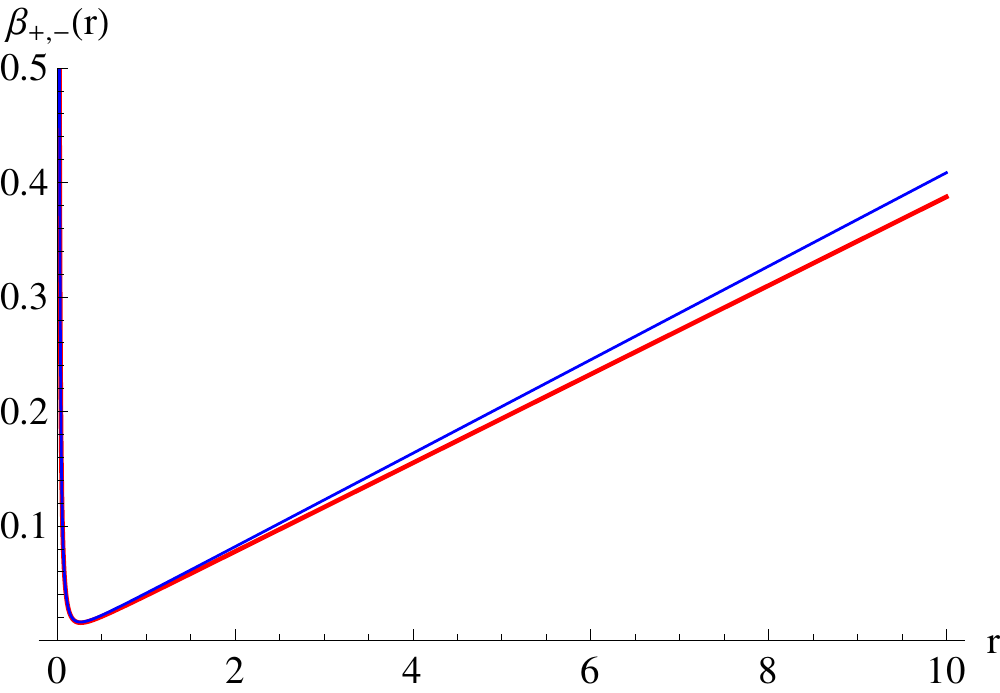}
\caption{\label{fig:Vr}Left: an example of the effective potential $V(r)$, where $\kappa = 10$, $M_{+} \simeq 7.71953$, $M_{-} = 0.9999999 M_{+}$, $\ell_{+} = 10^{6}$, $\ell_{-} = \kappa M_{-}$, and $\sigma \simeq 0.00016781$. Right: corresponding extrinsic curvatures $\beta_{+}$ (red) and $\beta_{-}$ (blue).}
\includegraphics[scale=0.75]{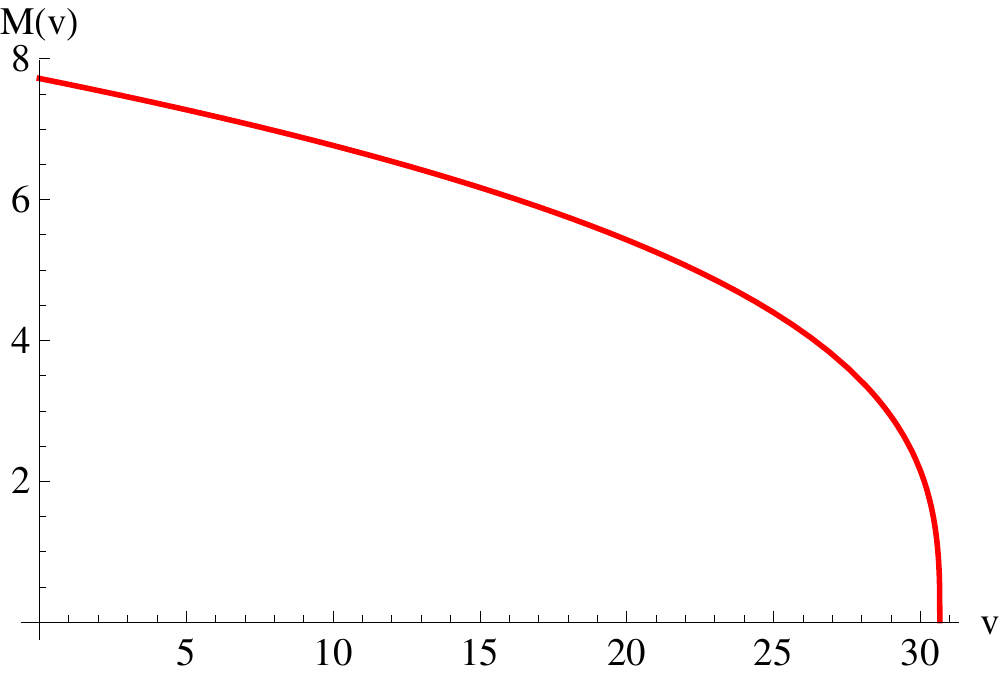}
\includegraphics[scale=0.75]{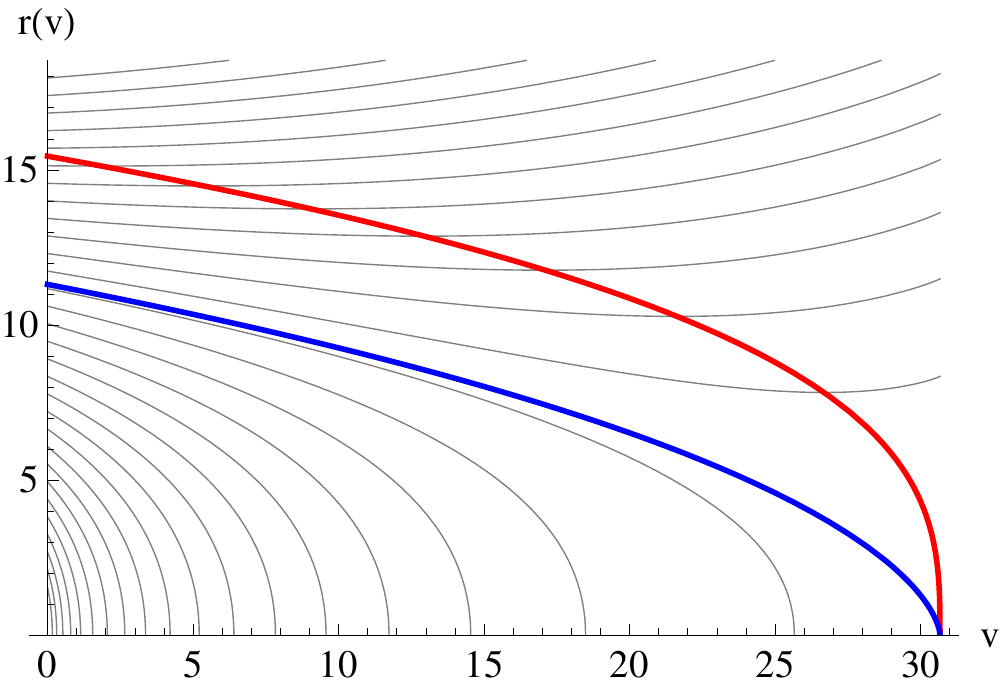}
\caption{\label{fig:mass}Left: $M(v)$ for the above case, where $\alpha N = 5$. Right: $r_{\mathrm{ApH}}$ (thick red), $r_{\mathrm{EH}}$ (thick blue), and some null geodesics (thin black). In this case, $r_{h} \simeq 14.8856$ and hence this is way outside the event horizon.}
\end{center}
\end{figure}

In the semi-classical limit, we can further estimate using analytic approximations \cite{COPSY}. After the evaporation, the event horizon $r_{\mathrm{EH}}$ is deviated from the apparent horizon $r_{\mathrm{ApH}}$ by the following factor:
\begin{eqnarray}
\frac{|r_{\mathrm{ApH}} - r_{\mathrm{EH}}|}{2M_{+}} \simeq \frac{4 \alpha N}{M_{+}^{2}} + \mathcal{O}\left(M_{+}^{-4} \right).
\end{eqnarray}
Then the condition for the left side to be naked is
\begin{eqnarray}
\frac{4 \alpha N}{M_{+}^{2}} - \frac{\delta r}{2M_{+}} > 0,
\end{eqnarray}
or
\begin{eqnarray}
N > \frac{M_{+}^{2}}{8 \alpha} \left( 2 - F(\kappa) \frac{M_{-}}{M_{+}} \right).
\end{eqnarray}
By further assuming $\Delta \ll M_{+}$ and $\kappa \gg 1$ limit,
\begin{eqnarray}
1 \lesssim \frac{M_{+}^{2}}{\alpha N} \ll \kappa^{2}
\end{eqnarray}
is required, where the first inequality is to be sure on the semi-classical limit (such that the formation time scale of a black hole $\sim M_{+}$ is shorter than the evaporation time scale $\sim M_{+}^{3}/\alpha N$ \cite{Dvali:2007hz}). Now, by choosing $\kappa$ and $\Delta$, we can find a valid limit that the left side can be naked to the right side\footnote{As we increase $\kappa$, at the same time $\ell_{-}$ should also be increased with a given $M_{-}$. Then perhaps, $M_{+}$ can be affected. However, we still have a freedom to choose $\kappa$. Hence, by tuning $\sigma$, one can adjust $\Delta / M_{+} \ll 1$.}. Indeed, it is possible to find such a set of parameters (FIGs.~\ref{fig:Vr} and \ref{fig:mass}).

\subsubsection{Non-adiabatic process} As discussed in \cite{COPSY}, there can be a non-adiabatic effect and this can be used to make the left side be naked. Although this non-adiabatic process is exponentially suppressed (e.g., \cite{Sasaki:2014spa,Chen:2015lbp,Chen:2015ibc}), if it is at once possible, then the horizon will be shifted approximately $\delta r \sim \Delta M$, where $\Delta M$ is the non-adiabatically emitted energy. Hence, it is possible to adjust $\delta r$ such that the signal from the left side can be seen by the right side observer.

\subsection{\label{sec:step5}Step 5: Communication via Einstein-Rosen bridge}

\begin{figure}
\begin{center}
\includegraphics[scale=1]{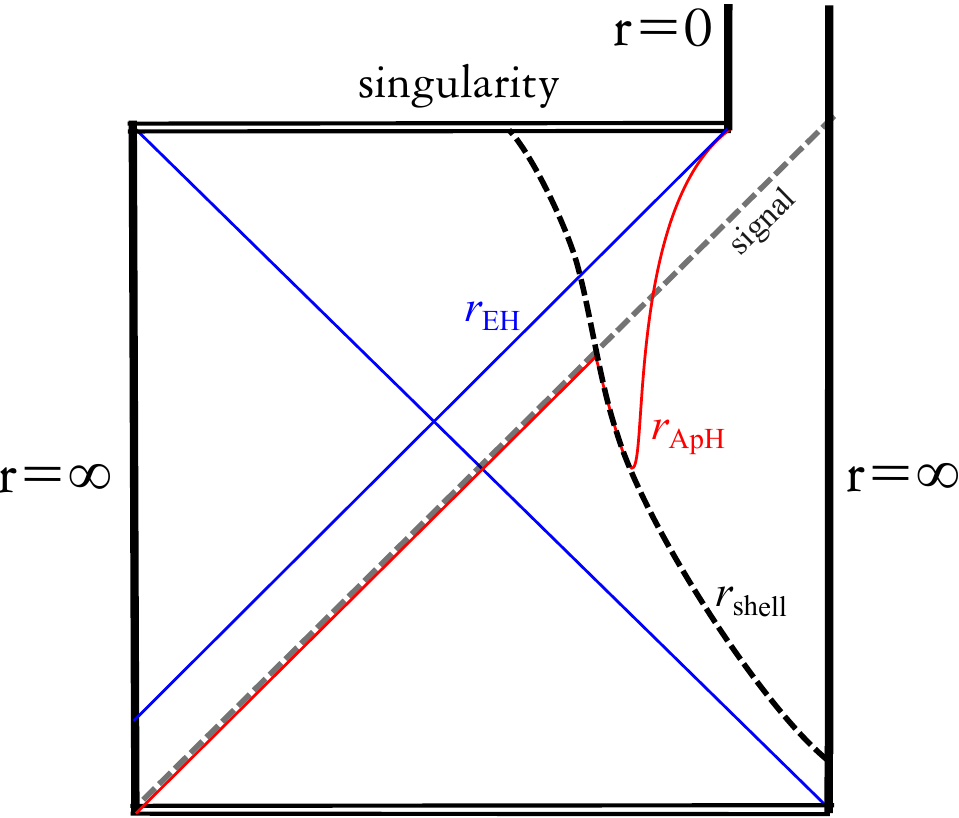}
\caption{\label{fig:horizons3}Then entire causal structure, including a past eternal black hole, a collapsing shell, and evaporation of the black hole $R$. Finally, a signal from the left side can be sent to the right side through the ER bridge.}
\end{center}
\end{figure}

If the horizon shrinks and the event horizon locates beyond the previous apparent horizon, then a signal from the left side can be transferred to the right side (FIG.~\ref{fig:horizons3}). Hence, if one sends a signal from the left side, then it can be naked in the right side of the universe.

\section{\label{sec:con}Conclusion and further debates}

In this paper, we invoked a thought experiment to argue against the ER=EPR conjecture. In order to do this, we considered maximally entangled systems that are separated by a distance. Due to quantum tunneling, the background AdS space tunneled toward a deeper vacuum. Each separated system is trapped by their respective nucleated bubbles, where one is expanding and the other is contracting. While each bubble is still large, we scramble the entangled systems and this will cause the formation of the ER bridge.

After the ER bridge is formed, the bubble for one side collapsed to the black hole. Hence, by choosing proper parameters, we can induce an evaporation of the black hole of only one side. Then, the apparent horizon as well as the true event horizon can shrink way smaller than the putative event horizon. Therefore, a signal from the left side can be transmitted to the right side of the ER bridge. This violates the principle of locality of EPR quantum states. This, in turn, serves as an evidence that the firewalls of eternal AdS black holes may also be naked. This may mean that firewalls can be naked not only for one-sided black holes \cite{Hwang:2012nn,Kim:2013fv,COPSY}, but also for two-sided ones.

Here we feel obliged to further compare the difference between our thought experiment and the ER=EPR conjecture. It is possible that the geometry of our thought experiment is not past-eternal, hence the lower left corner of the causal structure would be cutoff. Invoking the two-sided black holes to interpret the ER=EPR conjecture, where the two maximally entangled black holes are created at $t=0$, Maldacena and Susskind continue the black holes to the past of $t=0$ in the Penrose diagram \cite{MS}. In our counter example, we follow the same tactic, with the only but crucial difference that we consider in addition the back-reaction of the shells (see also \cite{Shenker:2013pqa,Shenker:2013yza,Shenker:2014cwa} for similar considerations but different conclusions). As a result, the duality between the EPR pairs and the ER bridge is broken.


The violation of the averaged null energy condition is a prerequisite to the traversability of wormholes \cite{Morris:1988tu}. In this regard, we note that there exist criticisms of theorems regarding the averaged null energy condition (e.g., \cite{Wall:2011hj})\footnote{We thank Juan Maldacena for bringing our attention to this issue \cite{private}.}. While the proof of the averaged null energy condition is not generic, it is possible to extend it to quite diverse theories. In particular, the averaged null energy condition around a bifurcation surface (e.g., the ER-like bridge) of a flat space is proven in \cite{Faulkner:2016mzt}, which can in principle be extended (but not yet proven) to curved spaces. Although there exist several proofs of the averaged null energy condition, so far it is still restricted to free or superrenormalizable theories, whereas in our case, we invoked a non-perturbative effect (a bubble) that requires renormalizable or non-renormalizable interactions. Therefore, even though there is a possibility to prove the averaged null energy condition for a bifurcation surface around the ER bridge, it is fair to say that an explicit proof is still lacking.

Even though the averaged energy condition is true, there are still some subtle issues to be reconsidered. The averaged null energy condition relies on some coarse-grained laws (e.g., the second law of thermodynamics \cite{C:2013uza}) or an averaged quantum energy-momentum tensor $\langle \hat{T}_{ab} \rangle$. These coarse-grained laws can in principle be violated \textit{locally}. As an example, let us include a non-perturbative history (a bubble) as we did in this paper. Then the wave function branches into approximately two histories, where one is without a bubble and the other is with a bubble. Since the probability of the latter history is highly suppressed, the contribution from this history to $\langle \hat{T}_{ab} \rangle$ is negligible and the averaged null energy condition is satisfied. However, if we observe a certain history locally, then it would not be so surprising that the ER bridge is traversable. Since a certain (even though highly suppressed) history allows a signal to transmit from one side to the other side, in terms of the entire wave function there should be a correlation between two sides that violates the locality. It is in this sense that our model serves as a counter-example to the ER=EPR conjecture.

There is actually a more severe criticism \cite{private}. That is, a moving shell can create particles due to the number $N$ matter fields and these particles can generate a positive in-going energy flux and may help to satisfy the averaged null energy condition. While this may indeed be possible, we note that the shell can in principle be constructed from a single scalar field and it does not necessarily couple to the other scalar fields. Hence, we believe that this would not be a critical problem to maintain our model.

There appeared several articles regarding the time-like nature of Einstein-Rosen bridge, e.g., Gao, Jafferis and Wall \cite{Gao:2016bin} and Maldacena, Stanford and Yang \cite{Maldacena:2017axo}, which deal with traversable wormholes by using the quantum energy-momentum tensor. According to the authors, due to the entanglement between the two sides of the ER bridge, the energy-momentum tensor could violate the averaged null energy condition. They argue that a signal transfer from one side to the other side of the ER bridge \textit{does not} violate the locality, but can instead be interpreted as a quantum teleportation. Regarding the traversability of the ER bridge, we agree that quantum effects (although our physical setting is different) may violate some of the energy conditions. However, in our case, the mechanism that opens the ER bridge is unrelated to the quantum entanglement between two sides, and hence there is no reason to connect the breaking of the ER bridge with quantum teleportation. Therefore, we conclude that our thought experiment is still valid as a counter-example to the ER=EPR conjecture.

If our thought experiment indeed holds, then the ER=EPR conjecture is necessarily restricted. Either this conjecture is simply wrong or it is restricted to only some specific models. This means that not all bulk geometries for maximally entangled pairs can be regarded as an ER bridge. We conclude that the ER=EPR conjecture cannot be a generic argument that substitutes the firewall.

There remain issues that require further investigations. What is the holographic dual of this communicating two-boundary system? What is the corresponding AdS/CFT dictionary? Evidently, we still have some distance to go toward the ultimate understanding of the information loss problem.

~\\


\section*{Acknowledgment}
The authors would like to thank Don N. Page for various comments and suggestions, and Juan M. Maldacena for valuable criticisms and communications. CHW would like to thank Ko-Wei Tseng and Ching-Che Yen for stimulated discussions.
DY was supported by Leung Center for Cosmology and Particle Astrophysics (LeCosPA) of
National Taiwan University (103R4000).


\newpage



\begin{thebibliography}{200}

\bibitem{Hawk} S. Hawking,
 \textit{Breakdown of Predictability in Gravitational Collapse},
  Phys. Rev. D {\bf 14}, 2460 (1976).

\bibitem{Chen:2014jwq} 
  P.~Chen, Y.~C.~Ong and D.~Yeom,
  \textit{Black Hole Remnants and the Information Loss Paradox},
  Phys.\ Rept.\  {\bf 603}, 1 (2015)
  [arXiv:1412.8366 [gr-qc]].

\bibitem{Ong:2016iwi} 
  Y.~C.~Ong and D.~Yeom,
  \textit{Black Hole Information Loss: Some Food for Thoughts},
  arXiv:1602.06600 [hep-th].

\bibitem{Malda} J. M. Maldacena,
 \textit{The Large N limit of superconformal field theories and supergravity},
  Adv. Theor. Math. Phys. {\bf 2}, 231 (1998) [hep-th/9711200].

\bibitem{Wit} E. Witten,
 \textit{Anti-de Sitter space and holography},
  Adv. Theor. Math. Phys. {\bf 2}, 253 (1998) [hep-th/9802150].

\bibitem{GKP} S. Gubser, I. R. Klebanov, and A. M. Polyakov,
 \textit{Gauge theory correlators from noncritical string theory},
  Phys. Lett. B {\bf 428}, 105 (1998) [hep-th/9802109].

\bibitem{LTU} L. Susskind, L. Thorlacius and J. Uglum,
 \textit{The Stretched Horizon and Black Hole Complementarity},
  Phys. Rev. D {\bf 48}, 3743 (1993) [hep-th/9306069].

\bibitem{Yeom:2008qw}
  D.~Yeom and H.~Zoe,
  \textit{Constructing a counterexample to the black hole complementarity},
  Phys.\ Rev.\  D {\bf 78}, 104008 (2008)
  [arXiv:0802.1625 [gr-qc]].

\bibitem{Hong:2008mw}
  S.~E.~Hong, D.~Hwang, E.~D.~Stewart and D.~Yeom,
  \textit{The causal structure of dynamical charged black holes},
  Class.\ Quant.\ Grav.\  {\bf 27}, 045014 (2010)
  [arXiv:0808.1709 [gr-qc]].

\bibitem{Yeom:2009zp}
  D.~Yeom and H.~Zoe,
  \textit{Semi-classical black holes with large N re-scaling and information loss problem},
  Int.\ J.\ Mod.\ Phys.\ A {\bf 26}, 3287 (2011)
  [arXiv:0907.0677 [hep-th]].

\bibitem{AMPS} A. Almheiri, D. Marolf, J. Polchinski and J. Sully,
 \textit{Black Holes: Complementarity of Firewalls?},
  JHEP {\bf 1302}, 062 (2013) [arXiv:1207.3123 [hep-th]].

\bibitem{AMPSS} A. Almheiri, D. Marolf, J. Polchinski, D. Standford and J. Sully,
 \textit{An Apologia for Firewalls},
  JHEP {\bf 1309}, 018 (2013) [arXiv:1304.6483 [hep-th]].

\bibitem{Page} D. N. Page,
 \textit{Average Entropy of a Subsystem},
  Phys. Rev. Lett. {\bf 71}, 1291 (1993) [gr-qc/9305007].

\bibitem{Page2} D. N. Page,
 \textit{Information in Black Hole Radiation},
  Phys. Rev. Lett. {\bf 71}, 3743 (1993) [hep-th/9306083].

\bibitem{Page3} D. N. Page,
 \textit{Time Dependence of Hawking Radiation},
  JCAP {\bf 1309}, 028 (2013) [arXiv:1301.4995 [hep-th]].

\bibitem{Hwang:2016otg} 
  J.~Hwang, D.~S.~Lee, D.~Nho, J.~Oh, H.~Park, D.~Yeom and H.~Zoe,
  \textit{Page curves for tripartite systems},
  arXiv:1608.03391 [hep-th].

\bibitem{MS} J. M. Maldacena and L. Susskind,
 \textit{Cool horizons for entangled black holes},
  Fortsch. Phys. {\bf 61}, 781 (2013) [arXiv:1306.0533 [hep-th]].

\bibitem{Hawking:1974sw} 
  S.~W.~Hawking,
  \textit{Particle Creation by Black Holes},
  Commun.\ Math.\ Phys.\  {\bf 43}, 199 (1975)
  Erratum: [Commun.\ Math.\ Phys.\  {\bf 46}, 206 (1976)].

\bibitem{Hayden:2007cs} 
  P.~Hayden and J.~Preskill,
  \textit{Black holes as mirrors: Quantum information in random subsystems},
  JHEP {\bf 0709}, 120 (2007)
  [arXiv:0708.4025 [hep-th]].
  
\bibitem{Hwang:2012nn} 
  D.~i.~Hwang, B.~H.~Lee and D.~Yeom,
  \textit{Is the firewall consistent?: Gedanken experiments on black hole complementarity and firewall proposal},
  JCAP {\bf 1301}, 005 (2013)
  [arXiv:1210.6733 [gr-qc]].

\bibitem{Kim:2013fv} 
  W.~Kim, B.~H.~Lee and D.~Yeom,
  \textit{Black hole complementarity and firewall in two dimensions},
  JHEP {\bf 1305}, 060 (2013)
  [arXiv:1301.5138 [gr-qc]].

\bibitem{COPSY} P. Chen, Y. C. Ong, D. N. Page, M. Sasaki and D. Yeom,
 \textit{Naked Black Hole Firewalls},
  Phys. Rev. Lett. {\bf 116}, 161304 (2016) [arXiv:1511.05695 [hep-th]].

\bibitem{Suss} L. Susskind,
 \textit{ER=EPR, GHZ and the Consistency of Quantum Measurements},
  Fortsch. Phys. {\bf 64}, 72 (2014) [arXiv:1412.8483 [hep-th]].

\bibitem{Suss2} L. Susskind,
 \textit{Copenhagen vs Everett, Teleportation, and ER=EPR},
  [arXiv:1604.02589 [hep-th]].

\bibitem{Israel:1976ur} 
  W.~Israel,
  \textit{Thermo field dynamics of black holes},
  Phys.\ Lett.\ A {\bf 57}, 107 (1976).

\bibitem{ER} A. Einstein and N. Rosen,
 \textit{The Particle Problem in the General Theory of Relativity},
  Phys. Rev. {\bf 128}, 919 (1935).

\bibitem{EPR} A. Einstein, B. Podolsky and N. Rosen,
  \textit{Can quantum mechanical description of physical reality be considered complete?},
   Phys. Rev. {\bf 47}, 777 (1935).

\bibitem{HMPS} I. Heemskerk, D. Marolf, J. Polchinski and J. Sully,
 \textit{Bulk and Transhorizon Measurements in AdS/CFT},
  JHEP {\bf 1210}, 165 (2012) [arXiv:1201.3664 [hep-th]].

\bibitem{MW} D. Marolf and A. C. Wall,
 \textit{Eternal Black Holes and Superselection in AdS/CFT},
  Class. Quant. Grav. {\bf 30} 025001 [arXiv:1210.3590 [hep-th]].

\bibitem{AC} S. G. Avery and B. D. Chowdhury,
 \textit{Firewalls in AdS/CFT},
  JHEP {\bf 1410}, 174 (2014) [arXiv:1302.5428 [hep-th]].

\bibitem{VR} M. Van Raamsdonk,
 \textit{Building up spacetime with quantum entanglement},
  Gen. Rel. Grav. {\bf 42}, 2323 (2010) [arXiv:1005.3035 [hep-th]].

\bibitem{Bell} J. Bell,
 \textit{On the Einstein-Podolsky-Rosen paradox},
  Physics {\bf 1}, 195-200 (1964).


\bibitem{FW} R. W. Fuller and J. A. Wheeler,
 \textit{Causality and Multiply Connected Space-Time},
  Phys. Rev. {\bf 128}, 919 (1962).

\bibitem{FSW} J. L. Friedman, K. Schleich and D. M. Witt,
 \textit{Topological censorship},
  Phys. Rev. Lett. {\bf 71}, 1486 (1993) [gr-qc/9305017].

\bibitem{GSWW} G. J. Galloway, K. Schleich, D. M. Witt and E. Woolgar,
 \textit{Topological censorship and higher genus black holes},
  Phys. Rev. D {\bf 60}, 104039 (1999) [gr-qc/9902061].

\bibitem{Kelly:2014mra} 
  W.~R.~Kelly and A.~C.~Wall,
  \textit{Holographic proof of the averaged null energy condition},
  Phys.\ Rev.\ D {\bf 90}, no. 10, 106003 (2014)
  Erratum: [Phys.\ Rev.\ D {\bf 91}, no. 6, 069902 (2015)]
  [arXiv:1408.3566 [gr-qc]].
  
\bibitem{Hartle:1976tp} 
  J.~B.~Hartle and S.~W.~Hawking,
  \textit{Path Integral Derivation of Black Hole Radiance},
  Phys.\ Rev.\ D {\bf 13}, 2188 (1976).
  
\bibitem{Malda2} J. M. Maldacena,
 \textit{Eternal black holes in anti-de Sitter},
  JHEP {\bf 0304}, 021 (2003) [hep-th/0106112].

\bibitem{Visser:1996iw} 
  M.~Visser,
  \textit{Gravitational vacuum polarization. 1: Energy conditions in the Hartle-Hawking vacuum},
  Phys.\ Rev.\ D {\bf 54}, 5103 (1996)
  [gr-qc/9604007].

\bibitem{Visser:1996iv} 
  M.~Visser,
  \textit{Gravitational vacuum polarization. 2: Energy conditions in the Boulware vacuum},
  Phys.\ Rev.\ D {\bf 54}, 5116 (1996)
  [gr-qc/9604008].

\bibitem{Coleman:1980aw} 
  S.~R.~Coleman and F.~De Luccia,
  \textit{Gravitational Effects on and of Vacuum Decay},
  Phys.\ Rev.\ D {\bf 21}, 3305 (1980).

\bibitem{Hawking:1982dh} 
  S.~W.~Hawking and D.~N.~Page,
  \textit{Thermodynamics of Black Holes in anti-De Sitter Space},
  Commun.\ Math.\ Phys.\  {\bf 87}, 577 (1983).

\bibitem{Blau:1986cw} 
  S.~K.~Blau, E.~I.~Guendelman and A.~H.~Guth,
  \textit{The Dynamics of False Vacuum Bubbles},
  Phys.\ Rev.\ D {\bf 35}, 1747 (1987).

\bibitem{Horowitz:2007pr} 
  G.~T.~Horowitz, J.~Orgera and J.~Polchinski,
  Phys.\ Rev.\ D {\bf 77}, 024004 (2008)
  doi:10.1103/PhysRevD.77.024004
  [arXiv:0709.4262 [hep-th]].
  
\bibitem{Harlow:2010az} 
  D.~Harlow,
  arXiv:1003.5909 [hep-th].

\bibitem{Israel:1966rt} 
  W.~Israel,
  \textit{Singular hypersurfaces and thin shells in general relativity},
  Nuovo Cim.\ B {\bf 44}, 1 (1966)
  [Erratum-ibid.\ B {\bf 48}, 463 (1967)].

\bibitem{Aguirre:2005xs} 
  A.~Aguirre and M.~C.~Johnson,
  \textit{Dynamics and instability of false vacuum bubbles},
  Phys.\ Rev.\ D {\bf 72}, 103525 (2005)
  [gr-qc/0508093].
  
\bibitem{Aguirre:2005nt} 
  A.~Aguirre and M.~C.~Johnson,
  \textit{Two tunnels to inflation},
  Phys.\ Rev.\ D {\bf 73}, 123529 (2006)
  [gr-qc/0512034].

\bibitem{Yeom:2016qec} 
  D.~Yeom,
  \textit{Information loss problem and roles of instantons},
  arXiv:1601.02366 [hep-th].

\bibitem{Vaidya:1951zz} 
  P.~Vaidya,
  \textit{The Gravitational Field of a Radiating Star},
  Proc.\ Indian Acad.\ Sci.\ A {\bf 33}, 264 (1951).


\bibitem{Dvali:2007hz} 
  G.~Dvali,
  \textit{Black Holes and Large N Species Solution to the Hierarchy Problem},
  Fortsch.\ Phys.\  {\bf 58}, 528 (2010)
  [arXiv:0706.2050 [hep-th]].

\bibitem{Sasaki:2014spa} 
  M.~Sasaki and D.~Yeom,
  \textit{Thin-shell bubbles and information loss problem in anti de Sitter background},
  JHEP {\bf 1412}, 155 (2014)
  [arXiv:1404.1565 [hep-th]].

\bibitem{Chen:2015lbp} 
  P.~Chen, G.~Dom{\`e}nech, M.~Sasaki and D.~Yeom,
  \textit{Stationary bubbles and their tunneling channels toward trivial geometry},
  JCAP {\bf 1604}, no. 04, 013 (2016)
  [arXiv:1512.00565 [hep-th]].

\bibitem{Chen:2015ibc} 
  P.~Chen, Y.~C.~Hu and D.~Yeom,
  \textit{Two interpretations of thin-shell instantons},
  Phys.\ Rev.\ D {\bf 94}, 024044 (2016)
  [arXiv:1512.03914 [hep-th]].

\bibitem{Shenker:2013pqa} 
  S.~H.~Shenker and D.~Stanford,
  \textit{Black holes and the butterfly effect},
  JHEP {\bf 1403}, 067 (2014)
  [arXiv:1306.0622 [hep-th]].
  
\bibitem{Shenker:2013yza} 
  S.~H.~Shenker and D.~Stanford,
  \textit{Multiple Shocks},
  JHEP {\bf 1412}, 046 (2014)
  [arXiv:1312.3296 [hep-th]].

\bibitem{Shenker:2014cwa} 
  S.~H.~Shenker and D.~Stanford,
  \textit{Stringy effects in scrambling},
  JHEP {\bf 1505}, 132 (2015)
  [arXiv:1412.6087 [hep-th]].

\bibitem{Morris:1988tu} 
  M.~S.~Morris, K.~S.~Thorne and U.~Yurtsever,
  \textit{Wormholes, Time Machines, and the Weak Energy Condition},
  Phys.\ Rev.\ Lett.\  {\bf 61}, 1446 (1988).

\bibitem{Wall:2011hj} 
  A.~C.~Wall,
  \textit{A proof of the generalized second law for rapidly changing fields and arbitrary horizon slices},
  Phys.\ Rev.\ D {\bf 85}, 104049 (2012)
  Erratum: [Phys.\ Rev.\ D {\bf 87}, no. 6, 069904 (2013)]
  [arXiv:1105.3445 [gr-qc]].

\bibitem{private} 
  J. M. Maldacena, private communications.

\bibitem{Faulkner:2016mzt} 
  T.~Faulkner, R.~G.~Leigh, O.~Parrikar and H.~Wang,
  \textit{Modular Hamiltonians for Deformed Half-Spaces and the Averaged Null Energy Condition},
  JHEP {\bf 1609}, 038 (2016)
  [arXiv:1605.08072 [hep-th]].

\bibitem{C:2013uza} 
  A.~C.~Wall,
  \textit{The Generalized Second Law implies a Quantum Singularity Theorem},
  Class.\ Quant.\ Grav.\  {\bf 30}, 165003 (2013)
  Erratum: [Class.\ Quant.\ Grav.\  {\bf 30}, 199501 (2013)]
  [arXiv:1010.5513 [gr-qc]].

\bibitem{Gao:2016bin} 
  P.~Gao, D.~L.~Jafferis and A.~Wall,
  \textit{Traversable Wormholes via a Double Trace Deformation},
  arXiv:1608.05687 [hep-th].

\bibitem{Maldacena:2017axo} 
  J.~Maldacena, D.~Stanford and Z.~Yang,
  \textit{Diving into traversable wormholes},
  Fortsch.\ Phys.\  {\bf 65}, no. 5, 1700034 (2017)
  [arXiv:1704.05333 [hep-th]].


%
%
%
%
%
%




%
%
%
%
%
%
%
%

%
%

\end{thebibliography}
\end{document}